\documentclass[11pt]{article}

\usepackage{amsmath,amsthm,amsfonts}
\usepackage{rotating} 
\usepackage{color,amsmath,amssymb,amsfonts}
\usepackage{epsf,epsfig,subfigure,psfrag,graphicx,afterpage}
\usepackage{url,lscape}
\usepackage[authoryear]{natbib}

\usepackage{multirow}
\usepackage{float}
\usepackage{rotating}
\usepackage{lscape}

\newcommand{\bftheta  }{\hbox{\boldmath$\theta$}}

\newcommand{\bfa}{\hbox{\boldmath$a$}}

\newcommand{\bfI}{\hbox{\boldmath$I$}}

\newcommand{\bfr}{\hbox{\boldmath$r$}}

\newcommand{\bfy}{\hbox{\boldmath$y$}}

\newcommand{\bfz}{\hbox{\boldmath$z$}}

\newcommand{\dint }{\displaystyle\int}

\newcommand{\bt}{{\scriptsize \bftheta}}

\begin{document}

\title{Zero Variance and Hamiltonian Monte Carlo
  Methods in GARCH Models}

\author{Rafael S. Paix\~ao$^{\rm a}$ and Ricardo S. Ehlers$^{\rm a}$\
\vspace{6pt}\\
$^{\rm a}${\em University of S\~ao Paulo, S\~ao Carlos, Brazil}
}

\date{}

\maketitle

\begin{abstract}

In this paper, we develop Bayesian Hamiltonian Monte Carlo methods for 
inference in asymmetric GARCH models under different distributions for the error term. We
implemented Zero-variance and Hamiltonian Monte Carlo schemes for
parameter estimation to try and reduce the standard errors of the estimates 
thus obtaing more efficient results at the price of a small extra computational cost.
\vskip .3cm


\end{abstract}

\section{Introduction}

ARCH and GARCH models first introduced by \cite{engle82} and
generalized by \cite{boll86} is certainly the most used class of
models to study the volatility of financial markets and have been
around for decades now (see for example \cite{terasvirta09} and \cite{tsay10}).
In terms of the statistical framework, these models provide motion dynamics for 
the dependency in the conditional time variation of the distributional parameters  
of the mean and variance, in an attempt to capture such phenomena as 
autocorrelation in returns and squared returns. Extensions to these models have 
included more sophisticated dynamics such as threshold models to capture the asymmetry
in the news impact. An example is the extended version of the GARCH model proposed in
\cite{gjr93} (GJR-GARCH hereafter). This composite model deals with possible asymmetries in the
volatility responses incorporating a leverage effect in the volatility equation. 

From a Bayesian perspective, computational methods based on Markov chain Monte Carlo (MCMC) 
have been utilized to address the complexity of these models. These are considered one of 
the most efficient estimation methods for this class of models. For a
survey on Bayesian estimation of GARCH models see for example \cite{ausin15}.

Zero-variance MCMC methods using control variates can dramatically
reduce the Monte Carlo variance of Bayesian estimates based on the
posterior expectation.
Recently, \cite{mira-etal2013} showed how ZV-MCMC can be applied
to estimate the parameters in a univariate GARCH(1,1) model with Gaussian
errors. Many empirical studies however, indicate that this model does
not account for the degree of kurtosis usually observed in most
financial time series. It is also known that the distribution of asset
returns exhibit the so called leverage effect, i.e. a negative past
return tends to increase the present volatility.
In this paper we discuss the use and efficiency of zero-variance MCMC
(ZV-MCMC) coupled with Hamiltonian Monte Carlo (HMC) methods
to estimate GJR-GARCH models. To increase flexibility we allow the error terms 
to follow normal, $t$-Student, Generalized
Error Distribution (GED) and the Generalized $t$ (GT) distributions (\cite{mcdonald-newey88}) with 
mean zero and unit variance.

The main contributions of the paper are to provide closed-form
analytic expressions necessary to implement the methods and assess
their statistical performances via simulation studies. We also
illustrate with real time series data. To our knowledge, such methods
have not been applied and studied before in the context of GJR-GARCH models.
All the computations in this paper were implemented using the 
{\tt C++} programming language through the {\tt Rcpp} and 
{\tt RcppArmadillo} which are available in the open-source statistical software
language and environment {\tt R} (\cite{r10}).

The remainder of the paper is organized as follows. Section \ref{meth}
describes the Bayesian model and the associated algorithms used for
inference. In Section \ref{sim} a simulation study is conducted to
illustrate the flexibity and computational gains in using the proposed
algorithms. Section \ref{res} presents a financial application using
index data and Section \ref{conclusion} concludes the paper.

\section{Methodology}\label{meth}

The so called GJR-GARCH model (\cite{gjr93}) is widely used to
describe leverage effects as it takes into account
possible asymmetries in individual assets volatilities. The model is
defined as, 
\begin{eqnarray*}
  x_y &=& y_y + \mu\\
  y_t &=& \epsilon_t \sqrt{h_t}~, ~\epsilon_t\sim D(0,1)\\
  h_t &=& \omega+\sum_{i=1}^p (\alpha_i + \phi_i I_{t-i})y_{t-i}^2+ \sum_{j=1}^q\beta_jh_{t-j},
\end{eqnarray*} 
where the $\epsilon_t$ are independent and identically distributed
error terms and
$D(0,1)$ denotes a distribution with mean zero and variance 1. Then,
given the past history at time $t$ denoted $F_{t-1}=\{y_{t-1},y_{t-2},\dots\}$,
$h_t$ is the conditional variance of $y_t$ written as an exact
function of the past. Also, $\omega>0$, $\alpha_i+\phi_i\ge 0, i=1,\dots,p$
and $\beta_j\ge 0, j=1,\dots,q$ define the positivity constraints and 
$\sum_{i=1}^p(\alpha_i+\phi_i/2)+\sum_{j=1}^q\beta_j < 1$, 
ensures covariance stationarity of $h_t$. In the variance
equation, the indicator function $I_{t-i}=1$ if $y_{t-i}\le 0$
and $I_{t-i}=0$ otherwise clearly induces a possibly asymmetric effect
of $y_{t-i}^2$ on $h_t$. Finally, it is clear that if $\phi_i=0, i=1,\dots,p$
this model reduces to the GARCH($p,q$) model with symmetric effects.

\subsection{Likelihood and Prior Distributions}\label{lp}

Following the Bayesian paradigm we need to complete the model
specification with appropriate prior distributions for the parameters.
The likelihood function,

\begin{eqnarray*}
f(y_t |h_t,\nu) = h_t ^{-1/2}\frac{1}{\sqrt{\pi (\nu-2)}}
\frac{\Gamma(\frac{\nu+1}{2})}{\Gamma(\frac{\nu}{2})}  
\left \{1 + \frac{y_t^2}{(\nu-2)h_t} \right\}^{-(\nu+1)/2}
\end{eqnarray*}
for $t$-Student errors with degrees of freedom parameter $\nu>2$ and,
\begin{eqnarray*}
f(y_t|h_t,\nu)=h_t ^{-1/2}
\frac{\nu}{\lambda 2^{1+ 1/\nu}\Gamma(1/\nu)}
\exp\left\{-\frac{1}{2\lambda^{\nu}}
\left|\frac{y_t}{\sqrt{h_t}}\right|^{\nu}\right\}
\end{eqnarray*}
for GED errors with shape parameter $\nu>0$ where 
$\lambda^2=2^{-2/\nu}\Gamma(1/\nu)/\Gamma(3/\nu)$. Since its kurtosis
is given by $\Gamma(1/\nu) \Gamma(5/\nu)/\Gamma(3/\nu)^2 -3$ then when 
$0<\nu < 2$ this distribution reproduces heavy-tails. Also, two
special cases of the GED are the standard normal distribution when
$\nu=2$ and the Laplace (or double exponential) distribution for
$\nu=1$. 

Finally, for the generalized $t$ errors,
\begin{eqnarray*}
f(y_t|h_t,\eta,\nu)=h_t ^{-1/2}
\frac{\eta\Gamma\left(\nu+\frac{1}{\eta}\right)}
{2\nu^{1/\eta}\Gamma\left(\frac{1}{\eta}\right)\Gamma\left(\nu\right)}
\left(1+\frac{1}{\nu}\left|\frac{x}{\sqrt{h_t}}\right|^{\eta}\right)^{-(\nu+1/\eta)},
\end{eqnarray*}
where $\eta>0$ and $\nu>0$ are two shape parameters. Larger
values of $\eta$ and $\nu$ yield a density with thinner tails than the
normal while smaller values are associated with thicker tailed
densities.

We propose the following prior distributions for the tail parameter $\nu$.
For GED errors, $\nu\sim N(0,\sigma^2_{\nu})$ truncated to $\nu > 0$
while for Student-$t$ errors we consider a $N(0,\sigma^2_{\nu})$ distribution
truncated to $\nu > 2$. For GT errors, $\eta\sim N(0,\sigma^2_{\eta})$ 
truncated to $\eta>1$ and $\nu\sim N(0,\sigma^2_{\nu})$ truncated to $\nu>1/\eta$.
For the parameters in the variance equation we also assign
truncated normal prior distributions, $\omega\sim N(0,\sigma^2_{\omega})$ truncated to $\omega>0$, 
$\alpha\sim N(0,\sigma^2_{\alpha})$ truncated to $0 < \alpha < 1$,
$\beta\sim  N(0,\sigma^2_{\beta})$ truncated to $0 < \beta < 1$,
$\phi \sim  N(0,\sigma^2_{\phi})$ truncated to $0< \phi < 2$. Finally,
we assign a normal distribution for mean parameter, $\mu\sim N(0,\sigma^2_{\mu})$.

Also, in order to employ the algorithms in this paper we need to implement
transformations of model parameters to the real line.
Here we propose to the following set of parametric transformations,
$\omega^{\ast} = \log(\omega) $, $ \alpha^{\ast} = \log\left( \frac{\alpha}{1 - \alpha} \right)$, 
$ \phi^{\ast} = \log\left( \frac{2\phi}{1 - \phi} \right)$, 
$ \beta^{\ast} = \log\left( \frac{\beta}{1 - \beta} \right)$.
For the parameters in the error distribution we take $\nu^{\ast} =\log(\nu + c)$ and 
$\eta^{\ast} = \log(\eta)$ where $c$ is the model constraint.

\subsection{Zero Variance Methods}

We assume that the aim is to estimate the expected value of a function
$f$ of the parameters $\bftheta$ with respect to their (unnormalized)
posterior distribution $\pi$, i.e. 
\begin{eqnarray*}
  \mu_f = E_{\pi}[f(\bftheta)]=
  \frac{\dint f(\bftheta)\pi(\bftheta) d\bftheta}{\dint\pi(\bftheta) d\bftheta}.
\end{eqnarray*}

\noindent Now, given a sample of size $N$,
$\bftheta^{(1)},\dots,\bftheta^{(N)}$, from the posterior
distribution, $\mu_f$ is estimated as $\hat{\mu}_f=\sum_{i=1}^N
f(\bftheta^{(i)})/N$. In order to reduce the Monte Carlo error of this
estimate the idea is to replace $f$ by a function $\tilde{f}$ which is
constructed such that $E_{\pi}(\tilde{f})=\mu_f$ but its variance is
much smaller. The zero-variance (ZV) principle was introduced in the
physics literature by
\cite{assaraf99} and \cite{assaraf03} and latter studied in
\cite{mira-etal2013} where rigorous conditions for unbiasedness and existence of
a central limit theorem (CLT) were derived. In what follows, we use
the re-normalized function $\tilde{f}$ defined as,
\begin{eqnarray}\label{tf}
  \tilde{f}(\bftheta)= f(\bftheta)+\frac{H\psi}{\sqrt{\pi(\bftheta)}}
\end{eqnarray}
where $H$ is a Hamiltonian operator and $\psi$ is an arbitrary trial
function. If Equation (\ref{tf}) satisfies $\mu_f=\mu_{\tilde{f}}$ then
$\tilde{f}$ can also be used to estimate the desired quantity
given a sample from the posterior distribution. So, in practice this
condition must be verified as it depends on the choices of $H$ and
$\psi$. Defining $\sigma_f$ as the standard deviation of $f$ with
respect to $\pi$, the optimal choice of $H$ and $\psi$ is obtained by
setting $\sigma_{\tilde{f}}=0$ (or equivalently $\tilde{f}=\mu_{f}$)
which in turn leads to the following fundamental equation,
\begin{eqnarray*}
  H\psi = -\sqrt{\pi(\bftheta)}[f(\bftheta)-\mu_{f}].
\end{eqnarray*}

Unfortunately, in most practical applications this cannot be solved
analytically and an approximate solution is proposed in
\cite{mira-etal2013} as follows. First specify the operator $H$ which is
required to be Hermitian and satisfy $H\sqrt{\pi}=0$. Then,
parameterize $\psi$ in terms of optimal parameters by minimizing
$\sigma_{\tilde{f}}$ and estimate these parameters from a short MCMC
simulation. Finally, a long MCMC simulation is performed from which
the posterior expectation is approximated using
$\hat{\mu}_{\tilde{f}}$ instead of $\hat{\mu}_{f}$. Here we adopt the
so called Schr\"odinger-type Hamiltonian operator proposed in
\cite{assaraf99}. For $\bftheta\in\mathbb{R}^d$,
\begin{eqnarray}\label{eq:h}
  H(\bftheta) = 
  -\frac{1}{2}\sum_{j=1}^d \frac{\partial^2}{\partial\theta_i^2}+V(\bftheta)
\end{eqnarray}
where
$V(\bftheta)=\frac{1}{2\sqrt{\pi(\bt)}}\Delta_{\bt}\sqrt{\pi(\bftheta)}$
and $\Delta_{\bt}$ denotes the Laplace operator of second derivatives
$\sum_{i=1}^d\frac{\partial^2}{\partial\theta_i^2}$. This choice of
operator guarantees the necessary condition $H\sqrt{\pi}=0$.

The trial function $\psi$ is typically chosen as
$\psi(\bftheta)=P(\bftheta)\sqrt{\pi(\bftheta)}$ where $P(\bftheta)$ is a
polynomial function. In principle, a higher variance reduction can be
achieved by simply using higher order polynomials. In practice
however, it has been reported in the literature that first and second
degree polynomials will provide considerable variance reduction (see
\cite{mira-etal2013}, \cite{papamg14} and \cite{oates16}).

This trial function is combined with
the Hamiltonian in (\ref{eq:h}) giving rise to the following
re-normalized function of $\bftheta$,
\begin{eqnarray*}
  \tilde{f}(\bftheta)= 
  f(\bftheta)-\frac{1}{2}\Delta_{\bt} P(\bftheta)+\nabla_{\bt} P(\bftheta) \bfz(\bftheta)
\end{eqnarray*}

\noindent where $\bfz(\bftheta)=-\frac{1}{2}\nabla_{\bt}\log\pi(\bftheta)$ and
$\nabla_{\bt}\log\pi(\bftheta)$ denotes the gradient vector of
$\log\pi(\bftheta)$ with respect to $\bftheta$. The elements in $\bfz$
are known as control variates.
In particular, for a first
degree polynomial, i.e $P(\bftheta)=\sum_{i=1}^d a_i\theta_i=\bfa'\bftheta$ it
follows that $\tilde{f}(\bftheta)= f(\bftheta) + \bfa'\bfz(\bftheta)$
and the coefficients $\bfa\in\mathbb{R}^d$ 
are chosen so as to minimize the variance of $\tilde{f}(\bftheta)$. It
can be shown that these coefficients are given by,
\begin{eqnarray}\label{eq:coeff}
\bfa = -Var[\bfz(\bftheta)]^{-1} Cov[f(\bftheta),\bfz(\bftheta)]
\end{eqnarray}
and are estimated from a short MCMC simulation which generates a
sample $\bftheta^{(1)},\dots,\bftheta^{(M)}$ from the posterior
distribution. Substituting the 
matrices in (\ref{eq:coeff}) by the sample variance
matrix of $\bfz(\bftheta^{(l)})$ and the sample covariance matrix
of $f(\bftheta^{(l)})$ and $\bfz(\bftheta^{(l)})$, $l=1,\dots,M$ 
we obtain an estimate $\hat{\bfa}$ of the coefficients. Once
$\hat{\bfa}$ is computed we can proceed and perform a much longer
simulation which generates a sample
$\bftheta^{(1)},\dots,\bftheta^{(N)}$ from the posterior distribution
of $\bftheta$. This gives rise to a sample of re-normalized functions,
\begin{eqnarray*}
  \tilde{f}(\bftheta^{(l)})= f(\bftheta^{(l)}) + \hat{\bfa}
  \bfz(\bftheta^{(l)}), ~l=1,\dots,N,
\end{eqnarray*}
and the posterior expectation of $\bftheta$ is estimated as 
$\hat{\mu}_{\tilde{f}}=\sum_{l=1}^N \tilde{f}(\bftheta^{(l)})/N$.

As was recently noticed by \cite{papamg14}, these ZV methods can be
embedded in sampling schemes that require the computation of gradients
of the log-target density in the first place. This is the case for so called
differential-geometric MCMC schemes like Hamiltonian Monte Carlo, 
Metropolis adjusted Langevin algorithm and their manifold extension
(see \cite{giro11} for details).

\subsection{Hamiltonian Monte Carlo}

The random walk Metropolis-Hastings algorithm is a commonly used
approach in the Bayesian literature for GARCH models. Hamiltonian
Monte Carlo (HMC) methods on the other hand combine Gibbs updates with
Metropolis updates and avoids the random walk behaviour which may lead
to inefficient exploration of the parameter space. This method
processes a new state by computing a trajectory obeying Hamiltonian
dynamics (\cite{nea2011}).

Consider a random vector $\boldsymbol\theta \in \mathbb{R}^k$ as position
variables (parameters) and  $\bfr \in \mathbb{R}^k$ an
independent auxiliary random vector with $\bfr\sim N_{k}(0,M)$. The
parameter space in then augmented by this auxiliary parameter $\bfr$ and
the joint probability density function of $(\bftheta,\bfr)$ is given by
\begin{equation}
\label{hmc:1}
\pi(\boldsymbol\theta , \bfr) \propto \exp(-H(\boldsymbol\theta, \bfr))
\end{equation}
where $H(\boldsymbol\theta, \bfr) = U(\boldsymbol\theta) + K(\bfr)$ 
is a Hamiltonian function,
$U(\bftheta)=-\log[\pi(\bftheta|y)\pi(\bftheta)]$ is refered to as the
\textit{potential energy} and $K(\bfr) = \bfr' M^{-1}\bfr/2$ is called
the \textit{kinetic energy}. A candidate value for $(\bftheta,bfr)$ is
generated in two steps. First, a value of $\bfr$ is simulated from a
normal distribution with mean zero and covariance matrix $M$
independently of $\theta$. Second, the joint system
$(\boldsymbol\theta,\bfr)$ made up of the current parameter value
and the new momentum $\bfr$ evolves through Hamiltonian dynamics
defined by the pair of differential equations,
$$
\frac{\partial \boldsymbol\theta}{\partial t}= 
\frac{H(\boldsymbol\theta, \bfr)}{\partial \bfr}= \frac{\partial K(\bfr)}{\partial \bfr};\qquad
\frac{\partial \bfr}{\partial t}= \frac{H(\boldsymbol\theta, \bfr)}{\partial \boldsymbol\theta}=
- \frac{\partial U(\boldsymbol\theta)}{\partial \boldsymbol\theta}.
$$
In practice however, these equations cannot be solved analytically and 
must be discretized using some small stepsize $\epsilon>0$. 
The leapfrog operator is typically used to solve Hamilton's equations 
(\cite{leimr04}) and works as follows, 
\begin{eqnarray*} 
r_{i}(t+\epsilon/2)&=&  
r_{i}(t) - (\epsilon/2)\frac{ \partial U(\boldsymbol\theta(t))}{\partial \theta_{i}}\\ 
\theta_{i}(t + \epsilon) &=&  
\theta_{i}(t + \epsilon/2) - \epsilon \frac{ \partial K(\bfr(t + \epsilon/2)) }{\partial r_{i}}\\
r_{i}(t+\epsilon) &=&
r_{i}(t+\epsilon/2)-(\epsilon/2)\frac{ \partial U(\boldsymbol\theta(t+\epsilon))}{\partial\theta_{i}}
\end{eqnarray*}

The resulting state at the end of $L$ repetitions of the above three
steps is a proposal $(\bftheta^*,\bfr^*)$ for the augmented parameter
vector. Since this discretization introduces an error, a Metropolis  
acceptance probability is employed to correct this bias and ensure
convergence to the invariant posterior distribution. The transition to a new proposed
value is accepted with probability, 
$$
\mathbb{P}((\bftheta,\bfr),(\bftheta^*,\bfr^*))=
\min\left\{1,\exp[H(\bftheta,\bfr)-H(\bftheta^*,\bfr^*)]\right\}
$$
and otherwise rejected.

\subsection{Influential Observations}\label{KL}

In this section, we propose to compute the Kullback-Leibler (K-L)
divergence to assess the potential influence of an observation in the posterior distribution
(see for example \cite{Hao-etal2016}).
The idea is to compare the posterior density
$p(\bftheta|\bfy)$ with a perturbed posterior density
$p_{\delta}(\bftheta|\bfy)$ by measuring a divergence between
$p(\bftheta|\bfy)$ and $p_{\delta}(\bftheta|\bfy)$. For a direct
comparison, the Kullback-Leibler divergence between $p(\bftheta|\bfy)$
and $p_{\delta}(\bftheta|\bfy)$ is given by,
\begin{equation}\label{KL1}
KL(p(\bftheta|\bfy),p_{\delta}(\bftheta|\bfy)) = 
\int p(\bftheta|\bfy)\log\left( \frac{p(\bftheta|\bfy)}{p_{\delta}(\bftheta|\bfy)}\right)d\bftheta.
\end{equation} 

We now define a general perturbation as the ratio of unnormalized posterior densities,
\begin{eqnarray}\label{eq:per1}
\delta(\bftheta,\bfy) &=&
\frac{p_{\delta}(\bfy|\bftheta)p_{\delta}(\bftheta)}{p(\bfy|\bftheta)p(\bftheta)}
\end{eqnarray}
from which the perturbed posterior density can be written as,
\begin{eqnarray*}
  p_{\delta}(\bftheta|\bfy)=
\frac{\delta(\bftheta,\bfy)p(\bftheta|\bfy)}
     {\int\delta(\bftheta,\bfy)p(\bftheta|\bfy)d\bftheta}.
\end{eqnarray*}
So in terms of perturbation, the divergence (\ref{KL1}) is written as,
\begin{equation}\label{KL2}
KL(p(\bftheta|\bfy),p_{\delta}(\bftheta|\bfy)) = 
\int p(\bftheta|\bfy)
\log\left(\frac{\delta(\bftheta,\bfy)}{\int\delta(\bftheta,\bfy)p(\bftheta|\bfy)d\bftheta}\right)^{-1}d\bftheta.
\end{equation} 

In particular, to assess the potential influence of observations only the likelihood function
is perturbed and the associated perturbation is given by,
\begin{eqnarray}\label{eq:per2}
\delta(\bftheta,\bfy)&=&
\frac{p(\bfy_{(i)}|\bftheta)}{p(\bfy|\bftheta)}=
\frac{\prod_{\stackrel{t=1}{t\ne i}}^n p(y_t|F_{t-1},\bftheta)}
     {\prod_{t=1}^n p(y_t|F_{t-1},\bftheta)},
\end{eqnarray}
where $\bfy_{(i)}$ denotes the series with the $i$th observation
deleted. In our GJR-GARCH(1,1) model however, for $t=i+1$ we need to redefine the conditional
variance and consequently the $(i+1)$th term of the likelihood function in
the numerator of (\ref{eq:per2}). It is not difficult to verify that,
$E(y_{i+1}|F_{i-1})=0$ and,
\begin{eqnarray*}
  h_{i+1}=Var(y_{i+1}|F_{i-1})&=&E(h_{i+1}\epsilon_{i+1}^2|F_{i-1})=E(h_{i+1}|F_{i-1})E(\epsilon_{i+1}^2|F_{i-1})\\
  &=& E(h_{i+1}|F_{i-1}) = \omega+[\alpha +\phi E(I_i|F_{i-1})]E(y_i^2|F_{i-1})+ \beta h_i\\
  &=& \omega+ (\alpha+\phi/2+\beta)h_i
\end{eqnarray*}
where we used the fact that $E(y_i^2|F_{i-1})=h_i$ and
$E(I_i|F_{i-1})=P(I_i=1|F_{i-1})=1/2$ for symmetric distributions.

Then, given a sample $\bftheta^{(1)},\dots,\bftheta^{(N)}$ from the posterior 
distribution of $\bftheta$ we obtain the following approximation for
the divergence of the $i$-th observation,
\begin{equation}\label{eq:Div_iw}
\widehat{KL}_{i} =\frac{1}{N}\sum_{l=1}^N
\left[-\log\left(\frac{\delta_i(\bftheta^{(l)},\bfy)}{\bar{\delta}_i}\right)\right], ~i=1,\dots,n.
\end{equation}
where
$\bar{\delta}_i=\frac{1}{N}\sum_{l=1}^N\delta_i(\bftheta^{(l)},\bfy)$
is the mean perturbation for the $i$-th observation. 

Higher values of this positive valued measure provide
indication for possibly influential observations. However, in a
similar vein as suggested in 
\cite{sanb16}, we propose to analyse these quantities relatively to
the others for each observation. This is accomplished by counting how many times
each value is greater than the others along the MCMC simulations. So,
given a sample from the posterior distribution we obtain samples of $KL$
divergences as, 
\begin{eqnarray*}
KL_{i}^{(l)} = 
-\log\left(\frac{\delta_i(\bftheta^{(l)},\bfy)}{\bar{\delta}_i}\right),
~i=1,\dots,n, ~l=1,\dots,N
\end{eqnarray*}
and for each $i$ we compute the following sample proportions,
\begin{eqnarray}\label{eq:po}
P_i = \frac{1}{N}\sum_{l=1}^N I\left(KL_{i}^{(l)} > \max_{j\ne i}KL_{i}^{(l)}\right), ~i=1,\dots,n.
\end{eqnarray}
Each of these proportions provides an estimate of the posterior probability of the $i$th
observation being influential. 

\section{A Simulation Study}\label{sim}


Here we concentrate on the performance of posterior means as parameter
estimators using HMC methods as opposed to traditional MCMC (Random
walk Metropolis) in the presence of variance reduction. We check the
effect of relaxing the normal errors assumption and model misspecification.

To generate the artificial data we considered GJR-GARCH(1,1) models with four 
distributions for the errors: Gaussian, GED with parameter $\nu=0.8$, 
Student $t$ with $\nu=8$ degrees of freedom and generalized $t$ distribution 
with shape parameters $\nu=4$ and $\eta=1$. For each model we generated $m=200$ 
time series with $n=200,500,1000,2000$ observations and parameters 
$\omega=0.05$, $\alpha=0.05$, $\phi=0.1$, $\beta=0.85$ and $\mu=0$. 
So, the simulated returns induce possible asymmetric effects on the volatilities.
Also, the persistence of the volatility processes, measured by 
$\alpha+\phi/2+\beta$ is 0.95 (a common finding when estimating such models).
We assume independent truncated normal prior distributions are as
described in Section \ref{lp} with large variances equal to 1000.

For the Metropolis-Hastings algorithm we proposed new parameter values from a 
multivariate normal distribution centered around the
current value with a variance-covariance proposal matrix which was calculated
from a pilot tunning procedure. Specifically, we construct a pilot sample of 2000 values from 
a Metropolis-Hastings with proposal variance-covariance matrix $\epsilon\bfI_k$.
A sample variance-covariance matrix $M$ is then calculated from the output of this pilot sample. 
The proposal variance-covariance matrix is $\epsilon M$ where $\epsilon$ is tuned 
so that the acceptance rates are close to 0.8.

For the HMC algorithm, we set $M=\bfI_k$, $L=20$ and chose a value for $\epsilon$ 
which lead to acceptance rates between 0.7 and 0.9. Two samples of size 2000 were 
generate discarding the first 1000 as burn-in. Then one sample was used to estimate 
the parameters in the zero variance method and the other one was used to estimate model parameters.

Tables \ref{E1}, \ref{E2}, \ref{E3} and \ref{E4} present the results from the simulations 
in terms of average bias and Monte Carlo standard errors.
Overall, the zero-variance methods provided a pretty small bias reduction 
(or no reduction at all) when used in tandem with HMC methods.

\begin{center}
  Tables \ref{E1}, \ref{E2}, \ref{E3} and \ref{E4} about here
\end{center}

The simulations also show that the zero-variance method is very efficient at 
reducing the standard error of the estimators, particularly when  a quadratic trial 
function is utilized. This is so for all sample sizes and error distributions. 
Take for example the standard error of the parameter $\omega$ in the GJR-GARCH(1,1) 
with normal errors which is 0.00015 for the HMC and is reduced to
0.00001 for the ZV-HMC-Q (a 15-fold reduction).

Finally, we note that the parameter $\nu$ in GJR-GARCH(1,1)-GT model
shows an odd behaviour when compared to the others. Its bias does not
decrease with the sample size and stays around 0.55. In other
simulations during this study we noticed that the parameters $\nu$ and
$\eta$ show a large posterior correlation.

\section{Empirical Results}\label{res}

\begin{figure}[h]\centering
\includegraphics{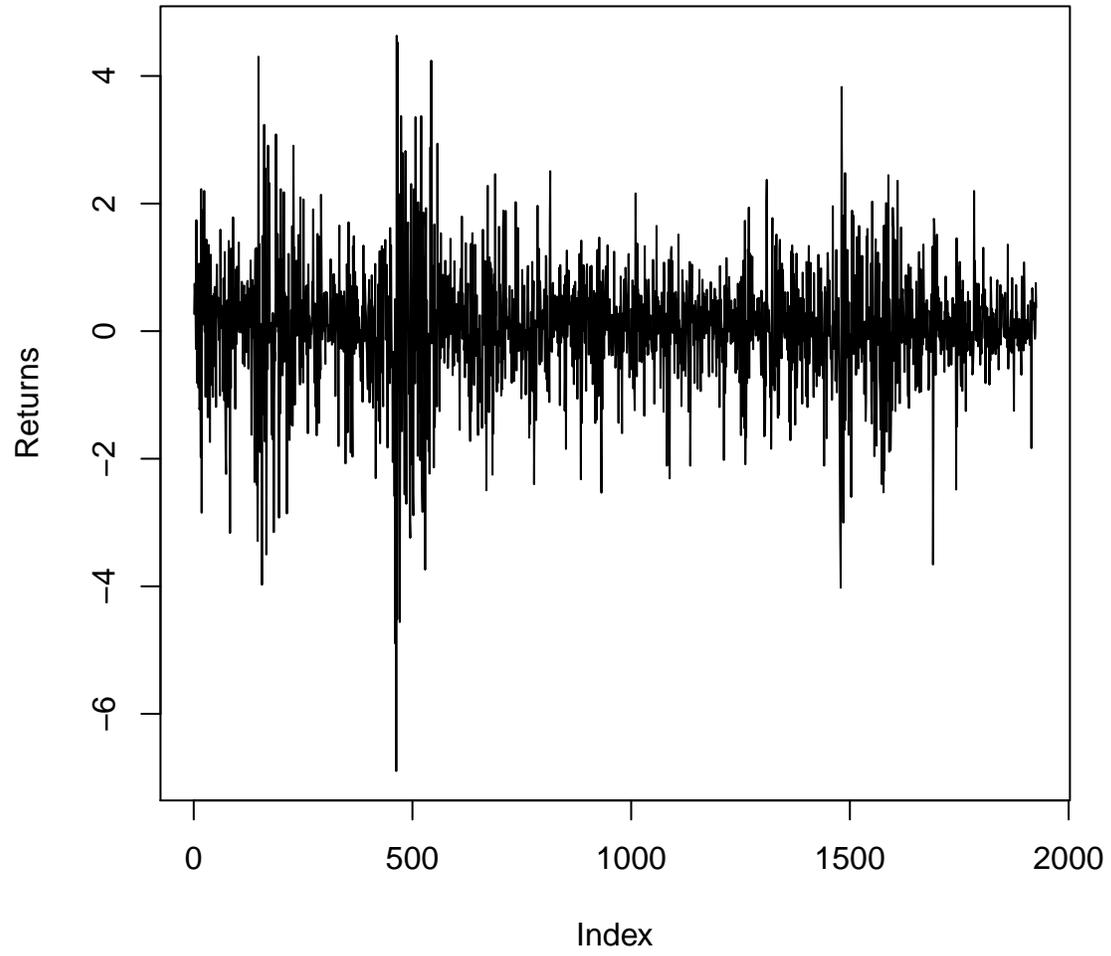}
\caption{S\&P500 time series returns from Oct 6, 2009 to June 2, 2017.}
\label{SP500}
\end{figure}

In this section, we illustrate the application of ZV, HMC and
influential observation methods by analysing some important stock
market indexes. We analysed the daily observations of the hundredfold
log-returns of daily indices of stock markets in Tokyo (SP\&500) from
Oct 6, 2009 until June 2, 2017, which leads to 1926 observations.
These data were obtained from Yahoo Finance and Figure \ref{SP500} shows
the time series returns. 

\begin{center}
Figure \ref{SP500} about here
\end{center}

Then, GJR-GARCH(1,1) models with the error terms
$\epsilon_t$ following Normal, GED with parameter $\nu$,
Student $t$ with $\nu$ degrees of freedom and Generalized $t$ distributions were estimated.
The prior distributions are again independent truncated normal distributions as described in 
Section \ref{lp} with variances all equal to 100.

For each model, we drew 10,000 samples of parameters using HMC
methods and discarded the first 5,000 as burn-in.
Table \ref{tab:DN} presents the point estimates (posterior means) and
the associated standard errors of the parameters for the
GJR-GARCH(1,1) model with normal, $t$, GED and generalized $t$ error distributions. We 
note that, the ZVHMC estimators have 
lower standard errors than the HMC estimates, especially when
the trial function is quadratic. In particular for the GJR-GARCH T
case, the simulated values of $\nu$ were highly correlated and
consequently the standard error is high. But applying the zero
variance principle we observe a large decrease in the standard errors.

\begin{figure}[h]
\centering
\includegraphics[scale=0.35]{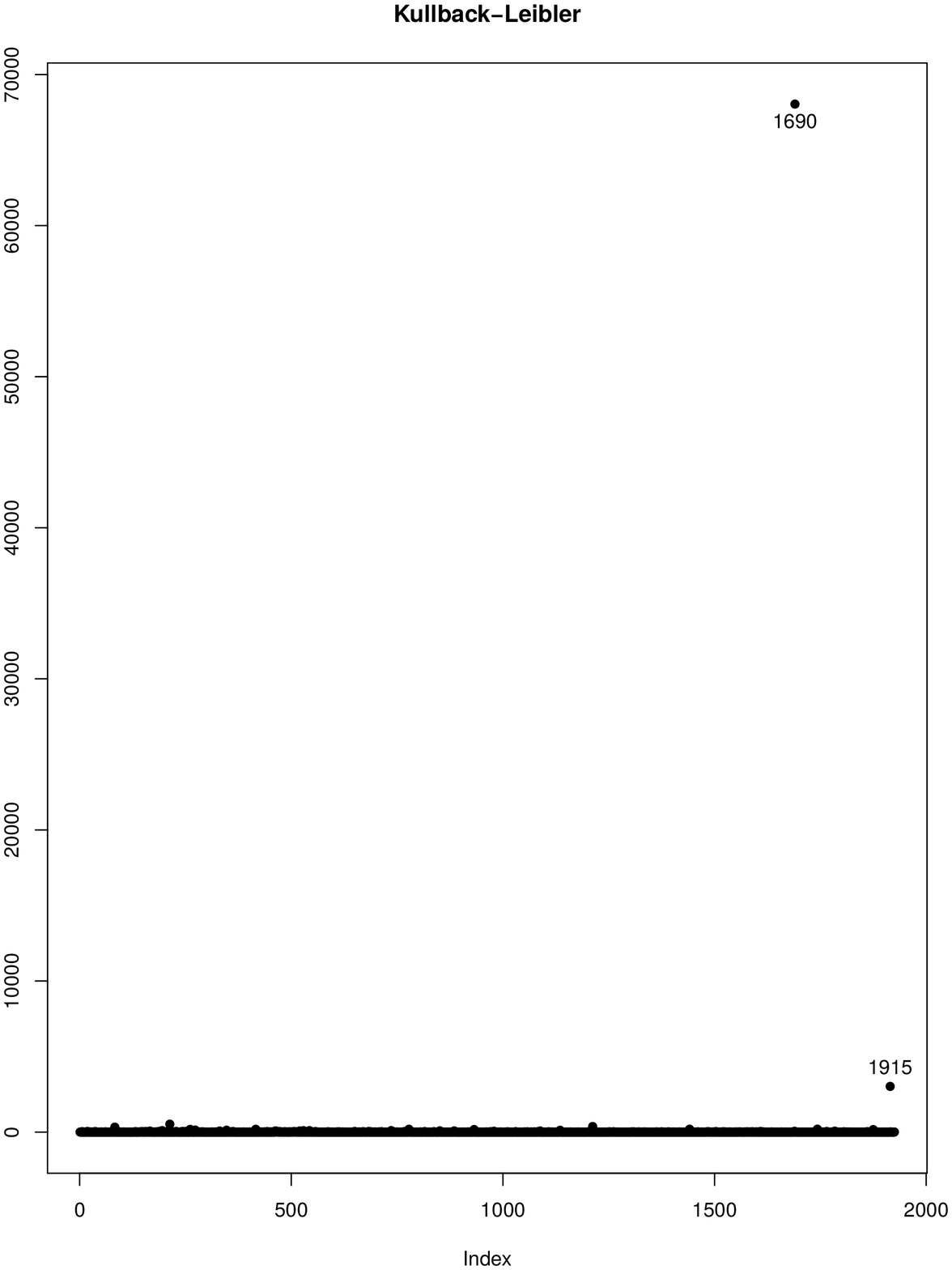}
\caption{Kullback-Leibler divergence for the GJR-GARCH(1,1)-GED model.}
\label{G:AI}
\end{figure}

The models were also compared in terms of the following information criteria, 
DIC (deviance information criterion), EAIC (expected AIC), EBIC
(expected BIC), DIC (deviance information criterion), WAIC (Watanabe
AIC) and LOOIC (leave-one-out information criterion). The results
appear in Table \ref{tab:crit}. From this table we notice that the
GJR-GARCH model with GED errors is clearly preferred in terms of all
the criteria considered. We then proceed with the influence analysis 
for this model using the Kullback-Leibler divergence as described in 
Section \ref{KL} to detect influential observations. The results are 
depicted in Figure \ref{G:AI}. We note that the point number 1690 (24 June 2016) 
is the most influential one and it represents a sharp decrease of 3.65\% in the index.
This is the date when the United Kingdom decided to leave the European Union.

\begin{center}
  Figure \ref{G:AI} about here
\end{center}

\section{Conclusion}\label{conclusion}

In this paper we discuss and compare the Bayesian estimation of GARCH
models 
with asymmetric effects and heavy tailed distributions under different
implementations of sampling the model parameters. Especifically, we
implemented Zero-variance (ZV) and Hamiltonian Monte Carlo schemes for
parameter estimation. Overall, the zero-variance method resulted in 
estimates with lower standard errors then HMC.
We also proposed tools to detect influential observations. 

These methods were assessed in both simulated data
and real time series of returns. The simulation studies provided
empirical evidence on the computational efficiency and variance
reduction. The additional computational cost involved is negligible compared to that
required to fit the model in the first place. 

\section*{Acknowledgements}

Ricardo Ehlers received
support from S\~ao Paulo Research Foundation (FAPESP) - Brazil, under
grant number 2016/21137-2.


\begin{thebibliography}{18}
\providecommand{\natexlab}[1]{#1}
\providecommand{\url}[1]{\texttt{#1}}
\expandafter\ifx\csname urlstyle\endcsname\relax
  \providecommand{\doi}[1]{doi: #1}\else
  \providecommand{\doi}{doi: \begingroup \urlstyle{rm}\Url}\fi

\bibitem[Assaraf and Caffarel(1999)]{assaraf99}
R.~Assaraf and M.~Caffarel.
\newblock {Zero-Variance principle for Monte Carlo algorithms}.
\newblock \emph{Physical Review Letters}, 83\penalty0 (23):\penalty0
  4682--4685, 1999.

\bibitem[Assaraf and Caffarel(2003)]{assaraf03}
R.~Assaraf and M.~Caffarel.
\newblock {Zero-variance zero-bias principle for observables in quantum Monte
  Carlo: application to forces}.
\newblock \emph{The Journal of Chemical Physics}, 119\penalty0 (20):\penalty0
  10,536--10,552, 2003.

\bibitem[Bollerslev(1986)]{boll86}
T.~Bollerslev.
\newblock Generalized autoregressive conditional heteroskedasticity.
\newblock \emph{Journal of Econometrics}, 31:\penalty0 307--327, 1986.

\bibitem[Engle(1982)]{engle82}
R.~F. Engle.
\newblock Autoregressive conditional heteroscedasticity with estimates of the
  variance of {U}nited {K}ingdom inflation.
\newblock \emph{Econometrica}, 50:\penalty0 987--1007, 1982.

\bibitem[Girolami and Calderhead(2011)]{giro11}
M.~Girolami and B.~Calderhead.
\newblock Riemann manifold {L}angevin and {H}amiltonian {M}onte {C}arlo
  methods.
\newblock \emph{Journal of the Royal Statistical Society B}, 73:\penalty0
  123--214, 2011.

\bibitem[Glosten et~al.(1993)Glosten, Jagannathan, and Runkle]{gjr93}
L.~R. Glosten, R.~Jagannathan, and D.~E. Runkle.
\newblock On the relation between the expected value and the volatility of the
  nominal excess returns on stocks.
\newblock \emph{The Journal of Finance}, 48:\penalty0 1791--1801, 1993.

\bibitem[Hao et~al.(2016)Hao, Lin, Wang, and Huang]{Hao-etal2016}
Hong-Xia Hao, Jin-Guan Lin, Hong-Xia Wang, and Xing-Fang Huang.
\newblock Bayesian case influence analysis for {GARCH} models based on
  {Kullback-Leibler} divergence.
\newblock \emph{Journal of the Korean Statistical Society}, 45\penalty0
  (4):\penalty0 595--609, 2016.

\bibitem[Leimkuhler and Reich(2004)]{leimr04}
Benedict Leimkuhler and Sebastian Reich.
\newblock {Simulating Hamiltonian dynamics, volume 14 of Cambridge monographs
  on applied and computational mathematics}.
\newblock \emph{Cambridge University Press, Cambridge}, 2:\penalty0 18, 2004.

\bibitem[McDonald and Newey(1988)]{mcdonald-newey88}
J.~B. McDonald and W.~K. Newey.
\newblock Partially adaptive estimation of regression models via the
  generalized $t$ distribution.
\newblock \emph{Econometric Theory}, 4:\penalty0 428--457, 1988.

\bibitem[Mira et~al.(2013)Mira, Solgi, and Imparat]{mira-etal2013}
A.~Mira, R.~Solgi, and D.~Imparat.
\newblock {Zero variance Markov chain Monte Carlo for Bayesian estimators}.
\newblock \emph{Statistics and Computing}, 23\penalty0 (5):\penalty0 653--662,
  2013.

\bibitem[Neal(2011)]{nea2011}
R.~M. Neal.
\newblock {MCMC} using {H}amiltonian dynamics.
\newblock In \emph{Handbook of Markov chain Monte Carlo}, pages 113--162. Boca
  Raton: Chapman and Hall-CRC Press, 2011.

\bibitem[Oates et~al.(2016)Oates, Girolami, and Chopin]{oates16}
C.~J. Oates, M.~Girolami, and N.~Chopin.
\newblock Control functionals for {M}onte {C}arlo integration.
\newblock \emph{Journal of the Royal Statistical Society: Series B (Statistical
  Methodology)}, 79\penalty0 (3):\penalty0 695--718, 2016.

\bibitem[Papamarkou et~al.(2014)Papamarkou, Mira, and Girolami]{papamg14}
T.~Papamarkou, A.~Mira, and M.~Girolami.
\newblock Zero variance differential geometric {M}arkov chain {M}onte {C}arlo
  algorithms.
\newblock \emph{Bayesian Analysis}, 9\penalty0 (1):\penalty0 97--128, 2014.

\bibitem[{R Development Core Team}(2015)]{r10}
{R Development Core Team}.
\newblock \emph{R: A language and environment for statistical computing}.
\newblock R Foundation for Statistical Computing, Vienna, Austria, 2015.

\bibitem[Santos and Bolfarine(2016)]{sanb16}
B.~Santos and H.~Bolfarine.
\newblock On {B}ayesian quantile regression and outliers.
\newblock \emph{ArXiv e-prints}, jan 2016.

\bibitem[Ter\"asvirta(2009)]{terasvirta09}
T.~Ter\"asvirta.
\newblock An introduction to univariate garch models.
\newblock In Davis R.A. Kreiss J.-P. Mikosch T.~(Eds.) Andersen, T.G., editor,
  \emph{Handbook of Financial Time Series}, pages 17--42. Springer, 2009.

\bibitem[Tsay(2010)]{tsay10}
R.~S. Tsay.
\newblock \emph{Analysis of Financial Time Series}.
\newblock John Wiley \& Sons, third edition, 2010.

\bibitem[Virbickait\.e et~al.(2015)Virbickait\.e, Ausin, and Galeano]{ausin15}
A.~Virbickait\.e, M.~C. Ausin, and P.~Galeano.
\newblock Bayesian inference methods for univariate and multivariate {GARCH}
  models: a survey.
\newblock \emph{Journal of Economic Surveys}, 29\penalty0 (1):\penalty0 76--96,
  2015.

\end{thebibliography}

\clearpage

\begin{landscape}
\begin{table}[h]
\centering
\begin{tabular}{r|r|rr|rr|rr|rr|rr|rr}
\hline 
\multicolumn{2}{c|}{ }	  &\multicolumn{2}{c|}{ HMC } &  \multicolumn{2}{c|}{ ZV-HMC-L}  &  \multicolumn{2}{c|}{ ZV-HMC-Q} &  
\multicolumn{2}{c|}{ RWM} &\multicolumn{2}{c|}{ ZV-RWM-L}  &  \multicolumn{2}{c}{ ZV-RWM-Q}  
\\ 
\hline 
N &	Par	& Bias& SE & Vis & SE &Vis &SE &Vis & SE &Vis &SE  & Vis & SE  \\
\hline
\multirow{5}{*}{200}
& $\omega$ & 0.04480& 0.00170 & 0.04480 &0.00147&  0.04475 &0.00078 & 0.04479 & 0.00072 & 0.03396 & 0.00053 &  0.03684 & 0.00043 \\
& $\alpha$ & 0.04973& 0.00459 & 0.05019 &0.00299&  0.05041 &0.00132 & 0.04972 & 0.00173 & 0.04331 & 0.00170 &  0.06101 & 0.00152 \\
& $\phi$   & 0.10084& 0.00754 & 0.10072 &0.00465&  0.10019 &0.00200 & 0.10084 & 0.00274 & 0.10723 & 0.00247 &  0.09675 & 0.00219 \\
& $\beta$  &-0.32998& 0.01156 &-0.33024 &0.00963& -0.33014 &0.00479 &-0.32996 & 0.00347 &-0.27652 & 0.00315 & -0.29693 & 0.00239 \\
& $\mu $   &-0.00459& 0.00084 &-0.00463 &0.00024& -0.00463 &0.00013 &-0.00459 & 0.00182 &-0.00565 & 0.00052 & -0.00832 & 0.00055 \\
\hline 
\multirow{5}{*}{500}	
&$\omega$ & 0.01149& 0.00071&  0.01148& 0.00051 & 0.01149& 0.00026&  0.01149& 0.00020&  0.00886& 0.00017&  0.01108& 0.00009\\
&$\alpha$ & 0.01750& 0.00347&  0.01747& 0.00235 & 0.01740& 0.00074&  0.01749& 0.00100&  0.01103& 0.00077&  0.01358& 0.00053\\
&$\phi$   & 0.03600& 0.00529&  0.03612& 0.00354 & 0.03607& 0.00112&  0.03600& 0.00120&  0.04505& 0.00118&  0.04486& 0.00070\\
&$\beta$  &-0.09709& 0.00633& -0.09695& 0.00398 &-0.09698& 0.00192& -0.09707& 0.00190& -0.08131& 0.00126& -0.09674& 0.00054\\
&$\mu$    &-0.00082& 0.00054& -0.00077& 0.00018 &-0.00077& 0.00007& -0.00081& 0.00119& -0.00175& 0.00019&  0.00202& 0.00017\\
\hline 
\multirow{5}{*}{1000}	
&$\omega$ & 0.00191& 0.00015&  0.00192& 0.00003&  0.00192& 0.00001&  0.00191& 0.00006&  0.00175& 0.00004&  0.00206& 0.00001\\
&$\alpha$ & 0.00712& 0.00167&  0.00717& 0.00087&  0.00707& 0.00019&  0.00712& 0.00045&  0.00346& 0.00024&  0.00449& 0.00006\\
&$\phi $  & 0.00457& 0.00274&  0.00457& 0.00147&  0.00474& 0.00038&  0.00459& 0.00050&  0.01044& 0.00038&  0.00937& 0.00009\\
&$\beta$  &-0.01863& 0.00172& -0.01873& 0.00032& -0.01871& 0.00008& -0.01863& 0.00084& -0.01761& 0.00031& -0.01958& 0.00004\\
&$\mu $   &-0.00082& 0.00022& -0.00081& 0.00008& -0.00082& 0.00003& -0.00083& 0.00060& -0.00130& 0.00005& -0.00003& 0.00002\\
\hline 
\multirow{5}{*}{2000}	
&$\omega$ & 0.00191& 0.00015&  0.00192& 0.00003 & 0.00192 &0.00001 & 0.00191 &0.00006 & 0.00175 &0.00004 & 0.00206 &<0.00001\\
&$\alpha$ & 0.00712& 0.00167&  0.00717& 0.00087 & 0.00707 &0.00019 & 0.00712 &0.00045 & 0.00346 &0.00024 & 0.00449 &0.00006\\
&$\phi  $ & 0.00457& 0.00274&  0.00457& 0.00147 & 0.00474 &0.00038 & 0.00459 &0.00050 & 0.01044 &0.00038 & 0.00937 &0.00009\\
&$\beta$  &-0.01863& 0.00172& -0.01873& 0.00032 &-0.01871 &0.00008 &-0.01863 &0.00084 &-0.01761 &0.00031 &-0.01958 &0.00004\\
&$\mu  $  &-0.00082& 0.00022& -0.00081& 0.00008 &-0.00082 &0.00003 &-0.00083 &0.00060 &-0.00130 &0.00005 &-0.00003 &0.00002\\
\hline 		
\end{tabular} 
\caption{Results from simulations for the GJR-GARCH(1,1)-NORMAL model.}
\label{E1}
\end{table}
\end{landscape}

\clearpage

\begin{landscape}
\begin{table}[h]
  \centering
  \begin{tabular}{r|r|rr|rr|rr|rr|rr|rr}
    \hline 
\multicolumn{2}{c|}{ }			& \multicolumn{2}{c|}{ HMC } &  \multicolumn{2}{c|}{ ZV-HMC-L}  &  \multicolumn{2}{c|}{ ZV-HMC-Q} &  
\multicolumn{2}{c|}{ RWM} &  \multicolumn{2}{c|}{ ZV-RWM-L}  &  \multicolumn{2}{c}{ ZV-RWM-Q}  
\\ 
\hline 
N &	Par	& Bias & SE & Bias & SE &Bias &SE &Bias & SE &Bias &SE  & Bias & SE  \\
\hline
\multirow{5}{*}{200}
&$\omega$&  0.05075& 0.00319&  0.05271& 0.00283&  0.05523& 0.00224&  0.05065& 0.00069&  0.03266& 0.00054&  0.03650& 0.00048\\
&$\alpha$&  0.06404& 0.00728&  0.06571& 0.00499&  0.06780& 0.00250&  0.06399& 0.00182&  0.04719& 0.00206&  0.05936& 0.00217\\
&$\phi  $&  0.13669& 0.01246&  0.13821& 0.00804&  0.13943& 0.00408&  0.13663& 0.00283&  0.13325& 0.00401&  0.14682& 0.00419\\
&$\beta$ & -0.35080& 0.01447& -0.35216& 0.01163& -0.35331& 0.00578& -0.35016& 0.00325& -0.28458& 0.00344& -0.29983& 0.00238\\
&$\mu  $ & -0.00548& 0.00079& -0.00540& 0.00025& -0.00534& 0.00014& -0.00552& 0.00163& -0.00742& 0.00058&  0.00058& 0.00075\\
&$\nu$   &  1.65575& 0.18313&  1.16832& 0.07342&  1.07581& 0.04158&  1.65374& 0.11214&  1.28561& 0.07894&  1.32999& 0.04894\\
\hline 
\multirow{5}{*}{500}	
&$\omega$&  0.01524& 0.00095&  0.01542 &0.00073&  0.01553& 0.00041&  0.01523& 0.00020&  0.01100& 0.00019&  0.01219& 0.00010\\
&$\alpha$&  0.02405& 0.00446&  0.02455 &0.00299&  0.02487& 0.00109&  0.02403& 0.00100&  0.01584& 0.00083&  0.01767& 0.00068\\
&$\phi$  &  0.04640& 0.00690&  0.04706 &0.00461&  0.04720& 0.00158&  0.04648& 0.00142&  0.05328& 0.00129&  0.06179& 0.00094\\
&$\beta$ & -0.13184& 0.00855& -0.13248 &0.00603& -0.13315& 0.00320& -0.13181& 0.00198& -0.10612& 0.00145& -0.11453& 0.00068\\
&$\mu$   & -0.00146& 0.00058& -0.00146 &0.00014& -0.00146& 0.00006& -0.00147& 0.00100& -0.00221& 0.00018&  0.00351& 0.00019\\
&$\nu$   &  1.80807& 0.14842&  1.52539 &0.04580&  1.51592& 0.01736&  1.80790& 0.07274&  1.12321& 0.04223&  1.69395& 0.01867\\
\hline 
\multirow{5}{*}{1000}
&$\omega$&  0.00760& 0.00062&  0.00770& 0.00038&  0.00778& 0.00020&  0.00760& 0.00013&  0.00645& 0.00011&  0.00811& 0.00003\\
&$\alpha$&  0.01633& 0.00264&  0.01671& 0.00180&  0.01679& 0.00058&  0.01627& 0.00044&  0.01473& 0.00037&  0.01745& 0.00020\\
&$\phi  $&  0.01324& 0.00376&  0.01380& 0.00255&  0.01385& 0.00085&  0.01334& 0.00078&  0.01686& 0.00061&  0.01118& 0.00028\\
&$\beta $& -0.06116& 0.00509& -0.06170& 0.00283& -0.06222& 0.00142& -0.06108& 0.00119& -0.05559& 0.00073& -0.06375& 0.00019\\
&$\mu   $& -0.00028& 0.00041& -0.00031& 0.00008& -0.00031& 0.00004& -0.00030& 0.00081& -0.00047& 0.00011& -0.00198& 0.00006\\
&$\nu   $&  1.27627& 0.10997&  1.10137& 0.03210&  1.09932& 0.00850&  1.27453& 0.05149&  0.60261& 0.02846&  1.28191& 0.00547\\
\multirow{5}{*}{2000}
&$\omega$&  0.00277& 0.00018&  0.00275& 0.00006&  0.00276& 0.00002&  0.00277& 0.00007&  0.00247& 0.00005&  0.00319& 0.00001\\
&$\alpha$&  0.00552& 0.00135&  0.00541& 0.00083&  0.00549& 0.00018&  0.00553& 0.00028&  0.00536& 0.00021&  0.00573& 0.00008\\
&$\phi $ &  0.01045& 0.00192&  0.01028& 0.00114&  0.01016& 0.00026&  0.01043& 0.00054&  0.01017& 0.00033&  0.01175& 0.00012\\
&$\beta$ & -0.02320& 0.00175& -0.02296& 0.00052& -0.02300& 0.00016& -0.02320& 0.00078& -0.02265& 0.00034& -0.02671& 0.00006\\
&$\mu  $ &  0.00009& 0.00025&  0.00009& 0.00004&  0.00009& 0.00001&  0.00010& 0.00060&  0.00005& 0.00005& -0.00001& 0.00002\\
&$\nu  $ &  0.62568& 0.05306&  0.63454& 0.01529&  0.63499& 0.00346&  0.62569& 0.03418&  0.37650& 0.01391&  0.65125& 0.00174\\
\hline 
\end{tabular} 
\caption{Results from simulations for the GJR-GARCH(1,1)-T model.}
\label{E2}
\end{table}
\end{landscape}

\clearpage

\begin{landscape}
\begin{table}[h]
  \centering
  \begin{tabular}{r|r|rr|rr|rr|rr|rr|rr}
    \hline 
\multicolumn{2}{c|}{ }		& \multicolumn{2}{c|}{ HMC } &  \multicolumn{2}{c|}{ ZV-HMC-L}  &  \multicolumn{2}{c|}{ ZV-HMC-Q} &  
\multicolumn{2}{c|}{ RWM} &  \multicolumn{2}{c|}{ ZV-RWM-L}  &  \multicolumn{2}{c}{ ZV-RWM-Q}  
\\ 
\hline 
N &	Par	& Bias & SE & Bias & SE &Bias &SE &Bias & SE &Bias &SE  & Bias & SE  \\
\hline
\multirow{5}{*}{200}
&$\omega$&  0.04351& 0.00218&  0.04319& 0.00185&  0.04293& 0.00092&  0.04351& 0.00131&  0.04280& 0.00109&  0.04838& 0.00086\\
&$\alpha$&  0.05606& 0.00638&  0.05583& 0.00404&  0.05592& 0.00177&  0.05619& 0.00291&  0.06365& 0.00357&  0.10990& 0.00262\\
&$\phi  $&  0.12018& 0.01058&  0.12024& 0.00655&  0.12059& 0.00282&  0.12028& 0.00453&  0.19138& 0.00608&  0.26531& 0.00529\\
&$\beta $& -0.32138& 0.01459& -0.32042& 0.01172& -0.31849& 0.00549& -0.32151& 0.00447& -0.32651& 0.00439& -0.35433& 0.00355\\
&$\mu   $& -0.00536& 0.00080& -0.00562& 0.00027& -0.00555& 0.00015& -0.00537& 0.00164& -0.00926& 0.00101& -0.00961& 0.00241\\
&$\nu   $&  0.03255& 0.00342&  0.03285& 0.00230&  0.03280& 0.00104&  0.03268& 0.02298&  0.04755& 0.00951& -0.02442& 0.00745\\
\hline 
\multirow{5}{*}{500}	
&$\omega$&  0.01319& 0.00074&  0.01320& 0.00053&  0.01305& 0.00029&  0.01326& 0.00045&  0.02150& 0.00039&  0.02293& 0.00022\\
&$\alpha$&  0.02575& 0.00375&  0.02487& 0.00256&  0.02710& 0.00084&  0.02670& 0.00175&  0.05007& 0.00152&  0.08252& 0.00100\\
&$\phi  $&  0.04456& 0.00611&  0.04423& 0.00411&  0.04458& 0.00129&  0.04439& 0.00238&  0.08529& 0.00315&  0.13963& 0.00185\\
&$\beta $& -0.11438& 0.00666& -0.11400& 0.00423& -0.11380& 0.00217& -0.11519& 0.00284& -0.18299& 0.00220& -0.20120& 0.00095\\
&$\mu   $& -0.00096& 0.00059& -0.00093& 0.00016& -0.00084& 0.00006& -0.00095& 0.00096& -0.00005& 0.00034& -0.02020& 0.00058\\
&$\nu   $&  0.01762& 0.00283&  0.01768& 0.00073&  0.01757& 0.00031&  0.01742& 0.01429& -0.02632& 0.00322& -0.07287& 0.00154\\
\hline 
\multirow{5}{*}{1000}	
&$\omega$&  0.00508& 0.00060&  0.00515& 0.00038&  0.00509& 0.00021&  0.00508& 0.00034&  0.01486& 0.00020&  0.01842& 0.00007\\
&$\alpha$&  0.01343& 0.00247&  0.01328& 0.00176&  0.01322& 0.00048&  0.01343& 0.00087&  0.03267& 0.00059&  0.04156& 0.00019\\
&$\phi  $&  0.01498& 0.00362&  0.01552& 0.00244&  0.01419& 0.00079&  0.01498& 0.00156&  0.03621& 0.00087&  0.05068& 0.00037\\
&$\beta $& -0.04750& 0.00482& -0.04640& 0.00277& -0.04648& 0.00144& -0.04750& 0.00177& -0.11817& 0.00119& -0.14952& 0.00037\\
&$\mu   $&  0.00074& 0.00039&  0.00068& 0.00010&  0.00070& 0.00004&  0.00074& 0.00077&  0.00075& 0.00013& -0.01364& 0.00020\\
&$\nu   $&  0.00593& 0.00078&  0.00557& 0.00033&  0.00554& 0.00012&  0.00593& 0.01004& -0.02630& 0.00145& -0.02943& 0.00041\\
\multirow{5}{*}{2000}
&$\omega$&  0.00241& 0.00019&  0.00237& 0.00006&  0.00239& 0.00007&  0.00241& 0.00016&  0.00449& 0.00009&  0.00597& 0.00002\\
&$\alpha$&  0.00694& 0.00164&  0.00682& 0.00110&  0.00614& 0.00053&  0.00697& 0.00054&  0.01664& 0.00032&  0.02207& 0.00008\\
&$\phi  $&  0.00729& 0.00171&  0.00647& 0.00102&  0.00626& 0.00034&  0.00729& 0.00109&  0.01575& 0.00049&  0.02400& 0.00013\\
&$\beta $& -0.02141& 0.00174& -0.02144& 0.00050& -0.02164& 0.00053& -0.02150& 0.00111& -0.04658& 0.00054& -0.05325& 0.00010\\
&$\mu   $& -0.00073& 0.00022& -0.00063& 0.00004& -0.00066& 0.00002& -0.00073& 0.00054& -0.00044& 0.00006& -0.00992& 0.00007\\
&$\nu   $&  0.00360& 0.00186&  0.00273& 0.00016&  0.00271& 0.00011&  0.00360& 0.00686& -0.02111& 0.00070& -0.01191& 0.00013\\
\hline 		
\end{tabular} 
\caption{Results from simulations for the GJR-GARCH(1,1)-GED model.}
\label{E3}
\end{table}
\end{landscape}

\clearpage

\begin{landscape}
\begin{table}
\centering
\begin{tabular}{r|r|rr|rr|rr|rr|rr|rr}
\hline 
\multicolumn{2}{c|}{ }	& \multicolumn{2}{c|}{HMC } &  \multicolumn{2}{c|}{ZV-HMC-L}  &  \multicolumn{2}{c|}{ ZV-HMC-Q} &  
\multicolumn{2}{c|}{RWM}& \multicolumn{2}{c|}{ZV-RWM-L} & \multicolumn{2}{c}{ZV-RWM-Q}  
\\ 
\hline 
N &	Par	& Bias & SE & Bias & SE &Bias &SE &Bias & SE &Bias &SE  & Bias & SE  \\
\hline
\multirow{5}{*}{200}
&$\omega$&  0.07396& 0.00695&  0.07622& 0.00695&  0.09096& 0.00765&  0.07397& 0.00119&  0.05184& 0.00123&  0.07539& 0.00165\\
&$\alpha$&  0.07271& 0.00983&  0.07288& 0.00748&  0.07262& 0.00586&  0.07263& 0.00241&  0.05137& 0.00365&  0.06210& 0.00395\\
&$\phi  $&  0.14984& 0.01694&  0.14864& 0.01264&  0.14859& 0.00934&  0.15023& 0.00399&  0.13891& 0.00512&  0.14811& 0.00538\\
&$\beta$ & -0.33664& 0.01721& -0.33653& 0.01433& -0.33882& 0.00903& -0.33654& 0.00413& -0.21153& 0.00457& -0.19179& 0.00456\\
&$\mu $  & -0.00305& 0.00103& -0.00302& 0.00053& -0.00299& 0.00039& -0.00306& 0.00147& -0.00492& 0.00098& -0.00305& 0.00125\\
&$\nu $  & -0.37755& 0.25731& -0.12751& 0.20128&  0.16583& 0.15173& -0.38517& 0.06114& -1.44091& 0.05476& -2.26104& 0.05945\\
&$\eta$  &  1.07731& 0.20018&  0.99143& 0.18878&  0.84586& 0.13160&  1.07707& 0.02539&  1.39641& 0.02394&  1.97826& 0.03804\\
\hline 
\multirow{5}{*}{500}	
&$\omega$&  0.01564 & 0.00117 & 0.01543& 0.00091&  0.01534& 0.00058&  0.01566& 0.00028 & 0.00901& 0.00027&  0.00704& 0.00019\\
&$\alpha$&  0.02713 & 0.00465 & 0.02648& 0.00316&  0.02601& 0.00126&  0.02710& 0.00135 & 0.01687& 0.00125&  0.03427& 0.00107\\
&$\phi  $&  0.05104 & 0.00740 & 0.05051& 0.00505&  0.05073& 0.00201&  0.05106& 0.00198 & 0.05924& 0.00177&  0.05829& 0.00143\\
&$\beta$ & -0.12636 & 0.00895 &-0.12557& 0.00603& -0.12572& 0.00305& -0.12648& 0.00242 &-0.08391& 0.00193& -0.07867& 0.00117\\
&$\mu  $ & -0.00201 & 0.00058 &-0.00196& 0.00019& -0.00199& 0.00008& -0.00203& 0.00090 &-0.00273& 0.00026& -0.00397& 0.00024\\
&$\nu  $ & -0.49023 & 0.19777 &-0.28602& 0.11533& -0.13005& 0.04942& -0.48954& 0.04001 &-1.40750& 0.03362& -1.13249& 0.02161\\
&$\eta $ &  0.29408 & 0.03933 & 0.27286& 0.02951&  0.24491& 0.01725&  0.29363& 0.00417 & 0.30186& 0.00437&  0.24474& 0.00351\\
\hline 
\multirow{5}{*}{1000}	
&$\omega$ & 0.00486& 0.00032&  0.00479& 0.00014&  0.00480& 0.00007&  0.00485 &0.00013&  0.00361& 0.00010&  0.00218& 0.00006\\
&$\alpha$ & 0.01208& 0.00262&  0.01187& 0.00173&  0.01192& 0.00050&  0.01206 &0.00082&  0.00800& 0.00063&  0.01531& 0.00041\\
&$\phi $  & 0.02265& 0.00450&  0.02233& 0.00289&  0.02213& 0.00085&  0.02273 &0.00103&  0.02729& 0.00093&  0.03658& 0.00058\\
&$\beta$  &-0.04598& 0.00338& -0.04576& 0.00127& -0.04599& 0.00054& -0.04596 &0.00154& -0.03809& 0.00083& -0.03356& 0.00034\\
&$\mu  $  &-0.00017& 0.00040& -0.00013& 0.00011& -0.00012& 0.00004& -0.00017 &0.00064& -0.00092& 0.00012& -0.00055& 0.00009\\
&$\nu  $  &-0.59149& 0.13937& -0.48429& 0.07439& -0.41785& 0.02636& -0.59531 &0.02821& -1.12365& 0.01765& -0.81978& 0.00959\\
&$\eta $  & 0.15264& 0.01502&  0.14557& 0.00853&  0.13993& 0.00266&  0.15292 &0.00171&  0.17968& 0.00170&  0.12871& 0.00097\\
\hline 		
\multirow{5}{*}{2000}	
&$\omega$&  0.00201& 0.00016&  0.00198& 0.00004&  0.00198& 0.00001&  0.00201& 0.00008&  0.00163& 0.00006&  0.00140& 0.00002\\
&$\alpha$&  0.00690& 0.00176&  0.00680& 0.00096&  0.00685& 0.00024&  0.00689& 0.00061&  0.00189& 0.00035&  0.00796& 0.00016\\
&$\phi $&   0.00610& 0.00310&  0.00582& 0.00168&  0.00567& 0.00048&  0.00610& 0.00070&  0.01286& 0.00052&  0.01403& 0.00022\\
&$\beta$&  -0.01909& 0.00188& -0.01897& 0.00037& -0.01900& 0.00011& -0.01908& 0.00109& -0.01614& 0.00047& -0.01740& 0.00011\\
&$\mu  $&  -0.00038& 0.00033& -0.00037& 0.00007& -0.00036& 0.00002& -0.00038& 0.00046& -0.00084& 0.00006& -0.00055& 0.00003\\
&$\nu  $&  -0.62460& 0.10178& -0.57874& 0.05532& -0.56899& 0.01670& -0.62555& 0.02136& -0.99361& 0.01078& -0.73694& 0.00355\\
&$\eta $&   0.10003& 0.01100&  0.09713& 0.00644&  0.09576& 0.00161&  0.10006& 0.00083&  0.12222& 0.00074&  0.08149& 0.00026\\
\hline
\end{tabular} 
\caption{Results from simulations for the GJR-GARCH(1,1)-GT model.}
\label{E4}
\end{table}
\end{landscape}

\clearpage

\begin{table}[h]
\centering
\caption{Parameter estimates (posterior means) and standard errors for
  the GJR-GARCH with normal, $t$ and GED errors.}
\label{tab:DN}
\vskip .2cm
\begin{tabular}{llcccccc}
\hline
       &         & HMC   & SE  &   ZV-HMC-L   & SE  & ZV-HMC-Q & SE   \\
\hline    
       &$\omega$ &0.04289& 0.00005& 0.04282& 0.00001& 0.042807& $<$0.00001\\
       &$\alpha$ &0.00623& 0.00023& 0.00558& 0.00007& 0.005423&  0.00002\\
Normal &$\phi$   &0.26407& 0.00018& 0.26416& 0.00007& 0.264357&  0.00002\\
       &$\beta$  &0.81459& 0.00018& 0.81491& 0.00003& 0.815006&  0.00001\\
       &$\mu$    &0.03091& 0.00036& 0.03114& 0.00001& 0.031127& $<$0.00001\\
\hline
       &$\omega$ &0.03772& 0.00008& 0.03791& 0.00001& 0.03799& 0.00001\\
       &$\alpha$ &0.00728& 0.00014& 0.00682& 0.00008& 0.00635& 0.00007\\
$t$    &$\phi$   &0.32830& 0.00071& 0.32965& 0.00045& 0.33049& 0.00004\\
       &$\beta$  &0.79973& 0.00038& 0.79928& 0.00005& 0.79925& 0.00002\\
       &$\mu$    &0.05480& 0.00025& 0.05441& 0.00006& 0.05429& 0.00000\\
       &$\nu$    &5.84457& 0.00505& 5.83454& 0.00518& 5.84157& 0.00133\\
\hline
       &$\omega$ & 0.04085& 0.00013& 0.04079& 0.00002& 0.04081& 0.00002\\
       &$\alpha$ & 0.00719& 0.00011& 0.00705& 0.00006& 0.00682& 0.00004\\
GED    &$\phi$   & 0.30951& 0.00053& 0.30834& 0.00022& 0.30894& 0.00019\\ 
       &$\beta$  & 0.79975& 0.00045& 0.80000& 0.00005& 0.79997& 0.00002\\
       &$\mu$    & 0.04607& 0.00019& 0.04675& 0.00005& 0.04664& 0.00002\\
       &$\nu$    & 0.64646& 0.00021& 0.64610& 0.00006& 0.64622& 0.00005\\
\hline
       &$\omega$ &0.03833& 0.00006& 0.03787& 0.00003& 0.03788& 0.00002\\
       &$\alpha$ &0.00665& 0.00023& 0.00620& 0.00010& 0.00648& 0.00009\\ 
GT     &$\phi  $ &0.33283& 0.00025& 0.33236& 0.00024& 0.33171& 0.00014\\
       &$\beta $ &0.79881& 0.00027& 0.79945& 0.00007& 0.79943& 0.00002\\
       &$\mu   $ &0.05460& 0.00028& 0.05470& 0.00005& 0.05482& 0.00002\\ 
       &$\nu   $ &2.63386& 0.01477& 2.65813& 0.00773& 2.64935& 0.00243\\ 
       &$\eta  $ &1.04019& 0.00209& 1.03909& 0.00084& 1.03949& 0.00029\\
\hline
\end{tabular}
\end{table}

\clearpage

\begin{table}[h]
\centering
\caption{EAIC, EBIC, DIC, WAIC and LOOIC values for the estimated GJR-GARCH(1,1) models.}
\label{tab:crit}
\begin{tabular}{lccccc}
\hline
               & DIC & EAIC   & EBIC   & WAIC   & LOOIC    \\
\hline    
GJR-GARCH(1,1)-NORMAL& 4687.4 & 4691.5 & 4713.8 & 4689.9 & 4689.9\\
GJR-GARCH(1,1)-T     & 4594.8 & 4600.1 & 4627.9 & 4595.4 & 4595.4\\
GJR-GARCH(1,1)-GED   & 4592.4 & 4597.5 & 4625.3 & 4593.1 & 4593.1\\
GJR-GARCH(1,1)-GT    & 4596.2 & 4603.6 & 4637.0 & 4596.9 & 4590.9\\
\hline
\end{tabular}
\end{table}

\clearpage

\clearpage

\end{document}